# Management's perspective on critical success factors affecting mobile learning in higher education institutions – An empirical study

Muasaad Alrasheedi, Luiz Fernando Capretz and Arif Raza
Department of Electrical and Computer Engineering, Western University
1151 Richmond St, London, Ontario, N6A 3K7, Canada
{malrash, lcapretz, araza22}@uwo.ca

**Abstract**

Mobile learning (m-Learning) is considered to be one of the fastest growing learning platforms. The immense interest in m-Learning is attributed to the incredible rate of growth of mobile technology and its proliferation into every aspect of modern life. Despite this, m-Learning has not experienced a similar adoption rate in the education sector, chiefly higher education. Researchers have attempted to explain this anomaly by conducting several studies in the area. However, mostly the research in m-Learning is examined from the perspective of the students and educators. In this research, it is contended that there is a third important stakeholder group whose opinion is equally important in determining the success of m-Learning: the university management. Although diversified by nature, heads of departments, deans, and IT system administrators are nevertheless considered members of any university management. The results of the research show that university commitment to m-Learning, university learning practices, and change management practices were the factors critical to the success of m-Learning, from the university management perspective.

**Keywords:** mobile learning, higher education, university management, critical success factors

**Introduction**

Mobile phones have found use in almost every aspect of modern day human life.
The versatility of mobile phone usage is the reason behind the global acceptance of this technology. The use of mobile phones has also extended to the education sector resulting in the development of a host of m-Learning platforms using wireless technology and portable handheld devices to impart education. The educational systems has been shaped by existing and emerging technologies practices (Capuruço & Capretz, 2009). Technology in education is becoming mobile-based with ever increasing use of smartphones and tablets. Many tools are being introduced to make the best use of technology in education. For example, Learning Management System (LMS) is considered an effective tool, particularly in the context of students' participation and their enhanced engagement in learning process (Park, 2014). Students are able to make use of this tool for all sorts of their academic activities such as downloading lecture notes and uploading assignments. Similarly faculty members can make use of the tool for uploading lecture notes, grades, etc.

Zeng and Luyegu (2011) referred to a series of pilot projects where technical feasibility and pedagogic integrations with mainstream educational methods are tested. As a result, many schools and universities are now part of these projects. Furthermore, new technologies such as mobile technologies will increasingly be used in the digital future (Kek and Huijser, 2011). Learners at this age are also more receptive of newer technologies, both hardware and software, which is an additional benefit for m-Learning applications at the level of higher education (Tsai et al., 2005).

Several surveys conducted by researchers have shown that students are almost entirely in favour of adopting m-Learning at the university level (Alrasheedi, 2015). Students tend to believe that this would definitely enhance their learning experience.

According to 2014 EDUCAUSE Report, nearly 86 percent undergraduate students owned a smart phone, while nearly 47 percent had tablets (Dahlstrom and Bichsel, 2014). However, statistics regarding the use of mobiles in learning reveal low penetration with only 30 percent of instructors incorporating mobile learning into assignments, and nearly 55 percent actually ban or discourage use of mobile devices during the class (Dahlstrom and Bichsel, 2014). The obvious reason for this discrepancy between the interest of learners and the actual adoption rate of an m-Learning platform, in light of the rapid growth of technology, is that some critical success factors impacting the adoption rate have been left unexplored (Zeng & Luyegu, 2011).

It is true that students are the most important of the user groups and are the target focus as well, but they are by no means the only stakeholder groups involved in decision making. There is a second stakeholder-user group that is equally important – the instructors. A few researchers have also extended their research in this direction. In this group, the scepticism towards m-Learning platforms becomes more apparent (Alrasheedi, 2015).

On the basis of our literature review, it has been realized that there exists a third stakeholder group that is generally overlooked in m-Learning research – the university management (higher level management, department heads, deans, and IT system administrators). Although they are the smallest group, they serve as the primary decision makers for any major technology adoption and hence their opinions and concerns are very important. The purpose of this paper is to present the assessment of the critical success factors of m-Learning from the perspective of university management.

The structure of the paper is as follows. Next section presents the literature review where several relevant aspects related to m-Learning and perception have been discussed. This is followed by the research model and the hypotheses to be tested. Afterwards, the research methodology, the analysis of data comprising a correlation analysis, and a determination of regression equation are presented. After discussion of the results and the limitations of the present study, the final section presents the conclusion.

**Literature Review**

*Concept of m-Learning*

The one feature that sets m-Learning apart from all other learning platforms is mobility. The notion of mobility is not merely limited to physical motion; the term mobility actually refers to the ability of a learner, instructor, or administrative staff or manager to have access to relevant information regardless of the time or place of access. This feature is not achievable when using non-mobile devices, as the name suggests- (Andrews et al., 2010). However, the idea of anytime-anywhere learning is theoretical; in practice, the learning is limited from being truly universal by factors such as connectivity, safety restrictions, and even privacy constraints (Saccol et al., 2010). Advantages of m-Learning are, however, not limited to mobility. M-Learning also brings in the key feature of collaborative learning. While collaboration is not a feature unique to an m-Learning platform, with the use of mobile devices the network of learners is wider than ever before. Further, mobile devices also take the idea of collaboration actively out from a formal classroom environment, making learning a much more dynamic activity **(**Kukulska-Hulme & Taxler, 2007). Moreover, the current growths in technology and the ubiquitous ownership of sophisticated mobile devices lead us to determine that the experience developed by teaching in this innovative classroom could be successfully adapted to more accustomed classroom in the future where collaborative learning activities take place through mobile devices (Salter et al. 2013).

*Understanding the concept of m-Learning*

Because of the use of technology in imparting education as well as the remoteness and, hence, mobility factor, the scope of m-Learning is fluid. The rapid advancements in mobile phones with both mainstream and obscure technologies mean a continual addition of features on a single device. This does add to the versatility of a handset, but at the same time makes it difficult to group various mobile devices under a single definition umbrella. The growth of the Internet is a further complication, as it brings its own brand of design challenges and usage constraints (Hamm et al., 2013).

Because m-Learning is a technology-intensive learning platform and actively uses the Internet as well as advanced versions of portable computers, many researchers tend to equate m-Learning with e-Learning, considering the former to be the successor of the latter (Kok, 2011). The authors agree with the notion given by Chaka (2009) that m-Learning is an upshot of distance-learning or d-Learning and e-Learning. Mobile technology principles make it technically possible to allow a non-contact, remote education scheme as a mainstream learning platform (Chaka, 2009).

*Barriers to adoption of m-Learning*

As can be seen from the discussion above, m-Learning offers several advantages, some of which are unique to this platform. Interestingly, however, every single one of its features has a downside attached to it. For instance, while mobile technology offers the prospect of flexible learning, this is not only limited by technology constraints but also by the interest and diligence of learners (Kukulska-Hulme, 2005). Zeldenryk and Bradey (2013) observe that students prefer flexible learning environment. The university management not only needs to ensure that the quality of learning remains the same across multiple platforms, but also has to take care of specific m-Learning related challenges like security, privacy, upgrading the platform to match the rapid technological changes, and developing multi-device compliant platforms, to name a few. Additionally, the management has to ensure that incorporating all the above provisions is done in a cost-effective manner, preferably resulting in cost savings or increased revenues in the long term (Ally, 2009). The extensive diffusion of mobile and wireless technologies is definitely not uniform and independent of economic and cultural factors. In fact, this diffusion offers a chance to create education policies aimed at increasing use of mobile devices in education (Seta, 2014).

*Previous studies*

The discrepancy between the high proliferation rates of mobile technology and new mobile-phone technologies and the modest adoption rates of the m-Learning platform in the higher education sector, has been the source of much interest to researchers. Several universities were actually a part of pilot studies reviewing the factors affecting adoption and the success of m-Learning (Ally, 2009). While it must be noted that m-Learning is based on the active interaction between humans and machines. This means that factors such as user experience, the social aspect, technical competency, etc., must be assessed in different contexts. Because these factors vary further based on the purpose of usage, they have to be assessed from the perspectives of various user groups – learners, educators, and university management (Andrews et al., 2010).

Researchers have actively assessed the critical success factors from the perspective of students (Alrasheedi, & Capretz, 2014; Pollara, 2011). Additionally, some researchers have also researched the opinions of instructors (Alrasheedi, Capretz, & Raza, 2015; Pollara, 2011). While these research studies are much fewer in number, the area has been explored to some extent. Critical success factors from university management perspective, thus, need to be studied in more detail. There are significant barriers to the adoption of an m-Learning platform, and many require active participation and support from the university management. Hence, it is important to understand their views on the subject. This paper presents an assessment of critical success factors from the university management perspective.

*Organizational behavior and organizational management: literature review*

Literature review has been performed by researchers on organizational theories (Ahmed & Capretz, 2010), organizational management (Ahmed & Capretz, 2007), and process evaluation (Ahmed, Capretz, & Samarabandu, 2008). They conclude that there are six factors – organizational structure, organizational culture, organizational commitment, organizational learning, change management, and conflict management – that are the most critical factors to address when studying the organizational perspective. In this research the same factors have been adopted and applied in order to present a foundation for the university management perspective as independent factors presented in this work.

Organizational structure is described by Wilson and Rosenfeld (1990) as the well-known pattern of interactions among the parts of an organization, outlining communication in addition to control and authority. As reported by Chatman (1996) and Wilson (2001) the organizational culture is categorized as involving a set of shared values, beliefs, assumptions, and practices that form and guide the attitudes and behaviour of entities within the organization. Moreover, Rosen (1995) mentioned that the internal orientation of workers is constructed mainly on the culture, beliefs, ethics, and expectations of that organization's workers and, consequently, has the prospect of being one of the greatest influential factors in strategic management. Additionally, organizational commitment is a performance attitude that is associated with the level of staff member contribution and to the intention to stay with the organization and is, accordingly, obviously associated to job performance (Mathieu & Zajac, 1990). Furthermore, organizational commitment has been summarized by Crewson (1997) as being a mixture of three recognizable factors relating staff cooperation: firstly, a firm belief in and respect of the organization's goals and values; secondly, excitement to work strong for the organization; and thirdly, ambition to continue with the same organization. Organizational learning is defined by Marquardt and Reynolds (1994) as a practice by which individuals acquire new skills and knowledge that govern their behavior and activities.

Organizational change, as defined by Beckhard and Harris (1987), is considered to be an organization's drive from its current phase to a future or target phase. Additionally, Todd (1999) describes change management as a systematic method that present a conceptual framework that includes process, politics, people, and strategy. According to Cao, Clarke, and Lehaney (2000), organizational change illustrates the variety of an organization and demonstrates the combination of technical and human actions that have inter-related purposes within the organization. Finally, conflict management involves analytic processes, inter-personal types, negotiating strategies, and other involvements that are considered to avoid unnecessary conflict and lower or resolve excessive conflict (Kottler, 1996).

**Research Model and Hypothesis**

In this paper, a research model has been developed for assessing how and to what extent different factors affect the perception of university management regarding the success of m-Learning in tertiary educational institutions. The six organizational factors, derived from Ahmed, Capretz, and Sheikh (2007), have been applied to a literature review of organizational theories in addition to organizational management and behaviour, in order to evaluate the university management perspective. The factors and the relationship model are shown in Figure 1.

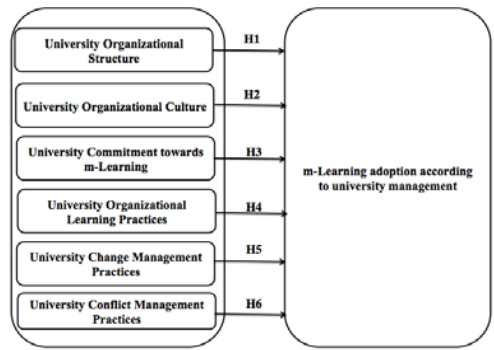

**Fig. 1.** Research model – Critical success factors affecting the success of m-Learning adoption from the perspective of university management.

The model proposed by Ahmed, Capretz, and Sheikh (2007) originally tested organizational factors that affect software product line performance. The rationale of borrowing the model to apply on m-learning is the fact that organizational factors influence decisions to implement any technology, as proved by Ahmed, Capretz, and Sheikh (2007). The model constitutes of three factors relating to organizational structure, and three relating to organizational behaviour. Using the same model, this study investigates the impact of University's organizational factors on the m-learning adoption.

To empirically investigate the research question, the six hypotheses have been derived as presented below:

Hypothesis 1.     The University Organizational Structure has a positive impact on m-Learning adoption, according to university management.

Hypothesis 2.      The University Organizational Culture has a positive impact on m-Learning adoption, according to university management.

Hypothesis 3.   The University Commitment towards m-Learning has a positive impact on m-Learning adoption, according to university management.

Hypothesis 4.   The University Organizational Learning Practices have a positive impact on m-Learning adoption, according to university management.

Hypothesis 5.   The University Change Management Practices have a positive impact on m-Learning adoption, according to university management.

Hypothesis 6.   The University Conflict Management Practices have a positive impact on m-Learning adoption.

University management is both the initial and final decision making authority to make policies and practices, both educational and IT policies. In general, academic management establishes educational policies and practices, whereas, technical policies and practices are governed by IT management. They are also responsible for platform upgrades, and, as system administrators, they form one of the user groups of the system. In this research, all six factors have been investigated that affect the overall attitude towards m-Learning adoption according to the perception of university management. To determine the management satisfaction levels a detailed survey (as illustrated in appendix1) has been conducted for assessing the factors affecting perception of university management regarding the success of the m-Learning platform.

Overall the objective of the research was to determine the answer to the following question:

"To what extent do the critical success factors have an impact on m-Learning adoption based on the perception of university management?"

**Research Methodology**

For collecting the data, an electronic questionnaire was sent to upper-level managerial staff (both academic and IT staff) working in various departments within five universities (Country name removed for the blind review). The staff was assured that their responses and identity would remain confidential and would not be disclosed. It was also explained to the staff that their primary responses were to be used only for this study. A total of 24 completed questionnaires were received from only three universities. The characteristics of users and their response pattern will be analyzed in the data analysis section below.

*Data collection and the measuring instrument*

As mentioned above, the present study involved getting responses from the university management level regarding their opinions on the issues affecting the success of m-Learning within their institution, and assessing their views on the subject. In order to determine this, an electronic survey questionnaire was sent to the management staff. In total 24 completed responses were received from management staff working at higher management levels from various departments within three universities. The analysis was performed using quantitative tools, specifically Minitab v.17 (Minitab, 2015).

*Reliability and validity of measuring instrument*

As the present survey was comprised of a set of demographic information, the questionnaire comprised a series of questions to determine the validity of the six hypotheses illustrated in Fig-1.

In each of the six hypotheses, the overall factor was determined using multi-item scales. Further, the dependent variable (m-Learning adoption) also comprised multi-item scales. Hence, in all these cases it was important to assess the reliability of the measurement scales. This was done to quantify the reproducibility of a measurement and was performed using an internal consistency analysis by calculating the Cronbach's alpha. The limits of satisfactory levels for this reliability coefficient have been determined by various researches. Most of the studies cite the work by Van de Ven and Ferry (2008) who considered that a coefficient of 0.55 and higher was satisfactory. Recent studies by researchers like Osterhof (2001), however, have increased the minimum satisfactory level of the reliability coefficient to be somewhat higher, 0.6. In our case, the reliability coefficient in all cases is >0.7, which means that the measuring instruments used are highly reliable.

Table 1. Cronbach's alpha for multi-measuring rating scales.

| Factors | Item Numbers | Cronbach's alpha | PCA Eigen Value |
|---|---|---|---|
| University organizational structure | H1 | 0.8089 | 1.051 |
| University organizational culture | H2 | 0.8922 | 1.038 |
| University commitment to m-Learning | H3 | 0.8436 | 1.456 |
| University organizational learning practices | H4 | 0.8849 | 1.402 |
| University change management practices | H5 | 0.9141 | 1.399 |
| University conflict management practices | H6 | 0.7299 | 1.315 |

The principal component analysis (PCA) was obtained for all six factors as reported in Table 1 (Kaiser, 1970). He argued that the Eigen Value was used as an indication point to identify the construct validity with PCA. The Eigen Value One criterion, which is known as the Kaiser Criterion (Kaiser, 1960; Stevens, 1986), was used which indicated that any component having an Eigen value greater than one should be retained. Eigen-value analysis revealed that all six variables form a single factor, as presented in Table 1. Consequently, based on our statistical analysis, the convergent validity of our measuring instrument can be considered as sufficient.

*Data analysis procedure*

For the present study, the data analysis process consisted of the following three steps. In the first step, a statistical check was performed to determine if there was a parametric correlation between the dependent variable and the independent variable. This was done to check if any of the critical success factors or hypotheses could be accepted statistically. In the second step, a non-parametric test was conducted between the dependent and independent variables. This was done in order to reduce the external validity threat (Raza, Capretz, & Ahmed, 2012). The third and final step of the statistical analysis comprised the regression analysis. This was done in order to determine the regression equation as discussed in following section, which gives the value and sign of the coefficients for each of the variables.

**Hypothesis tests and results**

*Hypothesis testing using parametric and non-parametric tests*

Before conducting the regression analysis, statistical tests were conducted to determine whether the relationships between the dependent variable and various independent variables were significant. This was done for each of the six hypotheses, using both parametric and non-parametric tests, by examining the Pearson and Spearman correlation coefficient. Further, it is a known fact that the lower the p-value the better chance there is of rejecting the null hypothesis and, hence, the result in terms of its statistical significance is more significant (Stigler, 2008). These two values were tested. The results are shown in Table 2 below.

Table 2. Hypothesis testing using parametric test and non-parametric statistical testing.

| Hypothesis | Critical Success Factors | Pearson Coefficient | Spearman Coefficient |
|---|---|---|---|
| H1 | University organizational structure | -0.051* | 0.127* |
| H2 | University organizational culture | -0.039* | 0.108* |
| H3 | University commitment towards m-Learning | 0.457** | 0.407** |
| H4 | University organizational learning practices | 0.402** | 0.457** |
| H5 | University change management practices | 0.399** | 0.420** |
| H6 | University conflict management practices | 0.316* | 0.238* |

** Significant at P < 0.05. * Insignificant at P > 0.05.

The results of the research show that the three factors – university commitment to m-Learning, university learning practices, and change management practices – were critical to the success of m-Learning from the university management perspective.

The Pearson correlation coefficient between the university commitment towards m-Learning and m-Learning adoption was positive (0.457) at P < 0.05, and, hence, hypothesis H3 is justified. For H4, the relationship between university organizational learning practices and the m-Learning adoption, the Pearson correlation coefficient, was 0.402 at P < 0.05, and, hence, it is found to be significant as well. Furthermore, hypothesis H5 was accepted based on the Pearson correlation coefficient of 0.399 at P < 0.05, which represents the relationship between the university change management practices and the m-Learning adoption according to the perception of university management. However, hypothesis H1, which denotes the relationship between the university organizational structure and m-Learning adoption, yields a Pearson correlation coefficient of (-0.051) at P = 0.27, and thus, this hypothesis is statistically insignificant; consequently, it was rejected. For H2, the relationship between the university organizational culture and the m-Learning adoption, the Pearson correlation coefficient, was (-0.039) at $P > 0.05$; hence, it was found to be insignificant and consequently, was rejected as well. Likewise, hypothesis H6 was rejected based on the Pearson correlation coefficient of 0.316 at $P > 0.05$, which represents the relationship between the university conflict management practices and the m-Learning adoption according to the perception of university management. Hence, as observed and reported, hypotheses H3, H4, and H5 were found to be statistically significant and were accepted, while H1, H2, and H6 were not supported and were, consequently, rejected.

In the second phase, non-parametric statistical testing was conducted by examining the Spearman correlation coefficient among the individual independent variables, the Critical Success Factors, and the dependent variable – m-Learning adoption according to the perception of university management, as displayed in Table 2.

Initially, the Spearman correlation coefficient between the university commitment towards m-Learning and the m-Learning adoption was found to be positive (0.407) at $P < 0.05$, and, hence, hypothesis H3 was justified. For hypothesis H4, which examined the relationship between university organizational learning practices and the m-Learning adoption, the Spearman correlation coefficient of 0.457 was observed at $P < 0.05$, and, hence, this hypothesis is significant. Moreover, hypothesis H5 was accepted based on the Spearman correlation coefficient of 0.420 at $P < 0.05$, demonstrating a statistically significant relationship between university change management practices and the m-Learning adoption as per the perception of university management. For hypothesis H1, which involves university organizational structure and the m-Learning adoption, the Spearman correlation coefficient of 0.127 was observed at $P > 0.05$. Since no significant relationship was found between the university organizational structure and the m-Learning adoption, H1 was rejected.

For H2, the relationship between the university organizational culture and the m-Learning adoption, the Spearman correlation coefficient was (0.108) at $P > 0.05$, and, hence, it was found to be insignificant; consequently, it was rejected too. Likewise, hypothesis H6 was rejected based on the Spearman correlation coefficient of 0.238 at $P > 0.05$, which represents the relationship between the university conflict management practices and the m-Learning adoption according to the perception of university management.

Hence, as observed and reported, H3, H4, and H5 were found to be statistically significant and were accepted, though H1, H2, and H6 were not supported and, hence, rejected in both parametric and non-parametric analysis.

*Testing of the research model using regression analysis*

The multiple linear regression equation of the model is as follows:

University management perception = $c_0 + c_1f_1 + c_2f_2 + c_3f_3 + c_4f_4 + c_5f_5 + c_6f_6$.

In the equation $c_0$, $c_1$, $c_2$, $c_3$, $c_4$, $c_5$, and $c_6$ are coefficients and $f_1$, $f_2$, $f_3$, $f_4$, $f_5$, and $f_6$ are the 6 independent variables.

In order to determine the coefficients of the equation above, a regression analysis was conducted. As can be seen from the model equation, all the critical success factors were assumed to have positive association with the m-Learning adoption as per the perception of university management by default. The results are given in Table 3 below.

The result of the regression analysis offer interesting insights into the model. First, not all the coefficients are positive. This means that three critical success factors – university organizational structure, university organizational culture, and university conflict management practices – all have negative association with university management perception. This deviates from the expected relationship.

The final regression equation is as follows:

$$m - Learning\ adoption\ as\ per\ University\ management\ perception$$
$$= 3.420 - 0.162(organizational\ structure) - 0.051(organizational\ culture)$$
$$+ 0.389(commitment) + 0.263(learning\ practices)$$
$$+ 0.036(change\ managment\ practices) - 0.334(conflict\ managment\ practices)$$

From the regression analysis, it is seen that the model accounts for only 37.01% variability in the dependent variable, i.e., m-Learning adoption.

Table 3. Multiple regression analysis of the research model.

| Critical Success Factor | Coefficient term | Coefficient value | t-value |
|---|---|---|---|
| University organizational structure | $f_1$ | -0.162 | -1.37 |

| University organizational culture | $f_2$ | -0.051 | -0.45 |
| University commitment towards m-Learning | $f_3$ | 0.389 | 1.66 |
| University organizational learning practices | $f_4$ | 0.263 | 1.71 |
| University change management practices | $f_5$ | 0.036 | 0.20 |
| University conflict management practices | $f_6$ | -0.334 | -1.13 |

**Discussion of results**

The data analysis section started with a detailed analysis of the demographic variables. This gives a snapshot of the population dynamics and characteristics. As the sample population of the study is only 24, it is not advisable to take this snapshot as a feature of management staff and their responses in a generic university setting. However, this can be taken as a case study. This is also one of the reasons demographic interrelationships have not been analyzed statistically as part of this study.

As all variables in the study comprised responses from multiple items in the survey, the reliability of the measuring instrument was tested first. This was done by determining the Cronbach's alpha for these multiple items. It was found that the value of Cronbach's alpha in most cases >0.7. As this is higher than the acceptable threshold of 0.6, using the average response for determining the individual variable coefficients could be done.

The next step was to determine if each of the independent-dependent variable pairs were correlated by finding out correlation coefficients. Both parametric and non-parametric studies were carried out to remove threats to external validity. It was found that the variables – university organizational structure, university organizational culture, and university conflict management practices – were not statistically significant as the p-values in each case was significantly >0.5.

Following this step, all six critical success factors were used for determining the regression model. It was found that the sign of the coefficients was negative for the three variables – university organizational structure, university organizational culture, and university conflict management practices. Interestingly, all other relationships were found to be positive though none of them had coefficients higher than 0.4. Also the highest correlation value was for university commitment to m-Learning followed by university learning practices. These also had the lowest p-values and significant t-values, showing that only these two relationships were worth investigating in future studies.

**Limitations of the study**

Empirical studies are subject to some limitations. In our study, the first limitation is the selection of independent factors. Only six independent variables were used to relate to the dependent variable of university management perspective. Although other factors might influence the university management perspective in addition to these six, the scope of this study was maintained within organizational management and behaviour as a base for the theoretical foundation. Despite the detailed nature of statistical analysis, this study has not explored the entire interrelationship between the demographic factors and the university management perception of the adoption of m-Learning within tertiary learning institutions. Some factors – such as gender, age group, management level, and even the department where the staff worked – might have an impact on the adoption of the new platform. The next step would have been the analysis of these variables. This means that based on the present results, a further study on how various demographic variables might have affected the perception of factors affecting m-Learning is redundant at this stage. The analysis can be a part of a future analysis, after more data is collected to see whether increasing the survey population changes the results. At the same time, future studies can also take into account more universities situated across different countries to improve the generalizability of the research.

**Conclusion**

The management level in a university is generally the ultimate authority regarding all decisions about if, when, and how a new learning platform has to be adopted. This research facilitates better understanding of the university management perspective about m-Learning adoption. Our main objective was to empirically investigate the effect of university factors on the adoption of m-Learning and find answers to the research question put forward in this investigation. Results of the research show that university commitment to m-Learning, university learning practices, and change management practices were the factors critical to the adoption of m-Learning from the university management perspective. A deeper understanding about the thought process of management staff is sure to help the adoption process of m-Learning. This was the core purpose behind conducting a study in this area.

The results of this investigation provide empirical evidence and further support the theoretical foundations that in order to have m-Learning within a university, the stated factors play an important role.

**Appendix1: Questionnaire on the university management perspective:**

**Part – I Opinions on the University's Organizational Structure**

Please rate the following statements according to your views on the university's current organizational structure.

1- Strongly Disagree, 2- Disagree, 3- Neither Agree or Disagree, 4-Agree, 5 - Strongly Agree

|  | 1 | 2 | 3 | 4 | 5 |
|---|---|---|---|---|---|
| 1. The roles and responsibilities of individuals and departments are clearly defined and documented. | [ ] | [ ] | [ ] | [ ] | [ ] |
| 2. The university's current organizational structure supports the m-Learning platform. | [ ] | [ ] | [ ] | [ ] | [ ] |
| 3. A strong and open communication channel exists between individuals/departments. | [ ] | [ ] | [ ] | [ ] | [ ] |
| 4. Employees are encouraged to work in interdisciplinary teams across department borders to share, disseminate, and acquire knowledge about the m-Learning platform. | [ ] | [ ] | [ ] | [ ] | [ ] |
| 5. All employees can directly communicate with the m-Learning support team | [ ] | [ ] | [ ] | [ ] | [ ] |
| 6. Cross-functional teams are established to monitor current m-Learning performance and to support management decision making. | [ ] | [ ] | [ ] | [ ] | [ ] |
| 7. The university's current strategic plan clearly defines how it will gain the technical capability to successfully adopt the m-Learning platform university-wide. | [ ] | [ ] | [ ] | [ ] | [ ] |

**Part – II Opinions on the University's Culture**

Please rate the following statements according to your views on the existing culture within the University

| | 1 | 2 | 3 | 4 | 5 |
|---|---|---|---|---|---|
| 1. The university's management welcomes new ideas to improve m-Learning acceptance. | [ ] | [ ] | [ ] | [ ] | [ ] |
| 2. New employees have difficulty in adapting to the university's working environment. | [ ] | [ ] | [ ] | [ ] | [ ] |
| 3. Employee opinions are asked and considered while implementing new ideas. | [ ] | [ ] | [ ] | [ ] | [ ] |
| 4. Employees are empowered to make appropriate decisions regarding job execution. | [ ] | [ ] | [ ] | [ ] | [ ] |
| 5. Employees are encouraged to work in interdisciplinary teams across department borders to share, disseminate, and acquire knowledge about the m-Learning platform. | [ ] | [ ] | [ ] | [ ] | [ ] |
| 6. Employees understand and are committed to the university's vision, values, and goals, chiefly in the area of m-Learning. | [ ] | [ ] | [ ] | [ ] | [ ] |
| 7. The university culture supports the reusability of software assets. | [ ] | [ ] | [ ] | [ ] | [ ] |
| 8. Higher management is generally viewed as approachable, supportive, and helpful. | [ ] | [ ] | [ ] | [ ] | [ ] |

**Part – III Opinions on the University's Commitment**

Please rate the following statements according to your views regarding the university's commitment

towards m-Learning

| | 1 | 2 | 3 | 4 | 5 |
|---|---|---|---|---|---|
| 1. The m-Learning platform is a clear part of the university's strategic vision. | [ ] | [ ] | [ ] | [ ] | [ ] |
| 2. University employees share a high degree of commitment to make the university's strategic vision a reality. | [ ] | [ ] | [ ] | [ ] | [ ] |
| 3. The employees feel a sense of ownership with the university rather than being just employees. | [ ] | [ ] | [ ] | [ ] | [ ] |
| 4. I would accept additional assignment in order to keep working with the university. | [ ] | [ ] | [ ] | [ ] | [ ] |
| 5. Over the last three years, on the whole, the university is steadily moving towards adopting an m-Learning platform as part of its strategic vision. | [ ] | [ ] | [ ] | [ ] | [ ] |
| 6. Employees consider m-Learning as a vital means to achieve the university's long-term goals. | [ ] | [ ] | [ ] | [ ] | [ ] |

**Part – IV Opinions on the University's Organizational Learning Practices**

Please rate the following statements according to your views regarding the university's organizational learning practices for employees.

|  | 1 | 2 | 3 | 4 | 5 |
|---|---|---|---|---|---|
| 1. Formal and informal learning programs are used to disseminate learning and knowledge within the university for its employees. | [ ] | [ ] | [ ] | [ ] | [ ] |
| 2. The necessary training has been provided to university employees on using the m-Learning platform. | [ ] | [ ] | [ ] | [ ] | [ ] |
| 3. The university is continuously in the process of learning from its experiences and lessons and avoids making the same mistake again and again. | [ ] | [ ] | [ ] | [ ] | [ ] |
| 4. Continuous monitoring and modification of the m-Learning platform has been taking place with respect to different comments and requirements. | [ ] | [ ] | [ ] | [ ] | [ ] |
| 5. Formal training sessions are regularly scheduled to train university staff on the m-Learning platform. | [ ] | [ ] | [ ] | [ ] | [ ] |
| 6. Employees share their experiences and knowledge with each other. | [ ] | [ ] | [ ] | [ ] | [ ] |

**Part – V Opinions on University's Change Management Practices**

Please rate the following statements, stating your views regarding the university's change management practices.

|  | 1 | 2 | 3 | 4 | 5 |
|---|---|---|---|---|---|
| 1. The university has a defined change management plan to adopt or switch to a new learning platform (e.g., m-Learning platform). | [ ] | [ ] | [ ] | [ ] | [ ] |
| 2. The change management program is well communicated to all the employees within the university. | [ ] | [ ] | [ ] | [ ] | [ ] |
| 3. The resistance to change to a newer platform (m-Learning) is gradually decreasing. | [ ] | [ ] | [ ] | [ ] | [ ] |
| 4. The changes in the organization with regarding to m-Learning platform adoption are well accepted by the employees. | [ ] | [ ] | [ ] | [ ] | [ ] |
| 5. The university regularly conducts reviews getting feedback from its employees on the m-Learning platform upgrades. | [ ] | [ ] | [ ] | [ ] | [ ] |
| 6. The university learns from the feedback and understands the impact of the newer platform on the organizational performance. | [ ] | [ ] | [ ] | [ ] | [ ] |

**Part – VI Opinions on University's Conflict Management Practices**

Please rate the following statements, stating your views regarding the university's conflict management practices

|  | 1 | 2 | 3 | 4 | 5 |
|---|---|---|---|---|---|
| 1. The university has a well-defined conflict management policy. | [ ] | [ ] | [ ] | [ ] | [ ] |
| 2. Management supports positive and constructive conflicts. | [ ] | [ ] | [ ] | [ ] | [ ] |
| 3. Personal conflicts are a major hurdle to the adoption of new practices and platforms. | [ ] | [ ] | [ ] | [ ] | [ ] |
| 4. Employees can successfully handle conflicts on their own. | [ ] | [ ] | [ ] | [ ] | [ ] |

**Part – VIII Opinions on the advantages of m-Learning platform**

Please rate the following statements, stating your views regarding the advantages of the m-Learning platform.

| | 1 | 2 | 3 | 4 | 5 |
|---|---|---|---|---|---|
| 1. The m-Learning platform has increased the capability of the university to manage students. | [ ] | [ ] | [ ] | [ ] | [ ] |
| 2. The m-Learning platform implementation has increased the student intake. | [ ] | [ ] | [ ] | [ ] | [ ] |